\documentclass[prd, letterpaper, 12pt,superscriptaddress,floatfix,preprintnumbers,altaffilletter, nofootinbib]{revtex4-1}
\usepackage{amsmath}
\usepackage{amssymb}
\usepackage{cancel}
\usepackage{color}
\usepackage{epsfig}
\usepackage[mathscr]{euscript}
\usepackage{graphicx}
\usepackage{longtable}
\usepackage{placeins}
\usepackage{rotating}
\usepackage{subfigure}
\usepackage{units}
\usepackage{url}
\usepackage{verbatim}
\usepackage{wasysym}
\usepackage{lipsum}
\usepackage{float}
\usepackage{mathtools}

\newcommand{\f}{\ensuremath{\mathrm{f}}}
\newcommand{\q}{\ensuremath{\mathrm{q}}}

\graphicspath {{}}

\begin{document}

\newcommand*{\rom}[1]{\uppercase\expandafter{\romannumeral #1\relax}}

\newcommand*\dif{\mathop{}\!\mathrm{d}}

\pagenumbering{arabic}

\title{A Carillon of Black Holes}

\author{Daniel George}
\email{dan.george@ligo.org}
\affiliation{Department of Physics, Pennsylvania State University, 104 Davey Lab, University Park, PA 16802, USA}

\author{Duncan Meacher}
\affiliation{Department of Physics, Pennsylvania State University, 104 Davey Lab, University Park, PA 16802, USA}

\author{Mark Ballora}
\affiliation{School of Music, Pennsylvania State University, University Park, PA 16802, USA}

\author{Chad Hanna}
\affiliation{Department of Physics, Pennsylvania State University, 104 Davey Lab, University Park, PA 16802, USA}
\affiliation{Department of Astronomy \& Astrophysics, Pennsylvania State University, 104 Davey Lab, University Park, PA 16802, USA}


\begin{abstract}
\vspace*{24pt}
Scientists collaborating internationally have developed a new way to learn
about our universe through gravitational waves, which are ripples in space-time
caused by the motion and vibration of celestial bodies.  By analogy,
gravitational waves are akin to the vibrations carried through the air as sound.
Quite remarkably, black holes, which are the densest objects in the universe
formed from dead stars can vibrate and emit gravitational waves at frequencies
that are within the range of human hearing once the gravitational waves are
detected and amplified by instruments such as the LIGO and Virgo gravitational
wave detectors.  In this work, we explore how to make musical instruments based on 
gravitational waves by mapping a different gravitational wave pattern to each of the 
88 keys of a piano, much like a carillon, which has its bells mapped to the batons of a 
carillon-keyboard. We rely on theoretical calculations for black hole 
vibrations to construct our digital black hole instruments. Our software and music 
samples are freely available to those who want to explore the music of gravitational waves.
\end{abstract}

\maketitle

\section{Introduction}
The universe contains a plethora of exotic objects that humans have been
studying through the medium of electromagnetic (EM) radiation, e.g., visible
light, for thousands of years. However, light is only one medium through which 
we are able to sense the Universe.  

In recent years, hearing has been regarded as another way of sensing cosmological 
phenomena. Cosmic microwave background radiation was first detected aurally from 
radio telescope signals, and it has been used in artistic outreach projects such as 
``Rhythms of the Universe," a short film by George Smoot and Mickey Hart~\cite{RhythmsUniverse}. 
Cosmologist Janna \cite{LevinTED} describes listening for gravitational waves (GWs) in her 
TED talk, which has the theme of ``space as a drum." Organizations such as the International Community for 
Auditory Display (www.icad.org) and Interactive Sonification (interactive-sonification.org) 
are dedicated to exploring the emerging use of sound in informatics, and researchers in a 
variety of fields are taking an interest in the potential of musically-based sound for new forms 
of data exploration and education/outreach.

Gravitational waves were first theorized in 1916 by Albert Einstein as a consequence of his 
theory of gravitation known as General Relativity (GR): he predicted that gravitational 
interactions were encoded as waves in the fabric of space-time. 
Although his theory has since been acknowledged and accepted, it took 100 years to obtain direct 
experimental evidence for GWs because the technology required to detect GWs involves measuring 
displacement much smaller than the diameter of a proton.
The ability to detect changes on this scale has only become feasible over the last decade
~\citep{harry2010advanced, aasi2015advanced, acernese2014advanced}.

On 14 September 2015, at 09: 50: 45 UTC, the two Laser Interferometer Gravitational-
Wave Observatory (LIGO) detectors~\citep{aasi2015advanced}, located at Livingston, LA, and Hanford, WA,
USA, made the first direct detection of GWs from two colliding
black holes (BHs)~\citep{abbott2016observation} that occurred some 1.3 billion light-years 
away.  With this detection, along with
several more in the years since~\citep{abbott2016gw151226, abbott2017gw170104,
abbott2017gw170814, abbott2017gw170817, abbott2017gw170608}, a new era of
gravitational wave astronomy as the Universe's soundtrack has expanded.

We are now able to literally `listen' to the GWs emitted from cataclysmic
astrophysical events in the Universe through the detections that LIGO makes. 
The present generation of GW 
detectors have a frequency response that lies within the human hearing range of
$20-20,000$~\footnote{LIGO can listen to sources within this range, but loses 
sensitivity after $8,000$ Hz} Hz.  A natural question to ask is, what do GW sources 
sound like? There are many GW signals that we could expect to 'hear'; 
chirping signals from colliding compact objects, burst signals from a supernova, or the 
background 'hum' from the birth of the Universe itself. 
Already, we have 'chirp' signal sounds from GWs emitted from the inspiral of binary-BH systems 
as BHs pick up speed right before merging. 
In this paper we consider the GWs emitted by the perturbation of a BH that could be caused through 
inspiral, impact, or other forms. We compare and contrast these sounds to those of traditional 
musical instruments such as drums and bells.  
We present a method to construct a digital instrument from a collection of BH spectra,
using results from a branch of physics known as BH perturbation theory.
We provide software for readers to construct their own BH instrument
mapped to the chromatic pitches found on a piano (github.com/georgedan1995/ACarillonOfBlackHoles). 

\parskip 18pt

%
\section{Gravitational waves and the sounds of a black hole}

Gravitational waves stretch and squeeze space-time in a pattern not dissimilar 
to ripples in water. 
As they pass a given area, the relative distance between two points within it changes 
as a function of time. The stretching and squeezing is characterized by the GW strain, 
$h(t)$, defined as
\begin{align}
h(t) &= \frac{\Delta L(t)}{L},
\end{align}
where $\Delta L(t)$ is the time-dependent change in distance between two points
separated by distance $L$.  Gravitational wave detectors such as LIGO and newly 
active detector Virgo measure $h(t)$ using laser interferometry and digitize 
the measured strain at a sample rate of 16,384 and 20,000 Hz respectively. The digitized 
strain time series can be normalized, filtered and encoded as a digital audio file.  
Thus, it is literally possible to listen to the ripples of the Universe.

Characterizing precisely what types of objects in the Universe produce GWs is an area 
of active research~\citep{cutler2002overview}, but here we will focus on GWs from 
perturbed BHs~\citep{ruffini1971introducing}.  Black holes are of particular 
interest since they are the predominant source detected by the LIGO and Virgo
GW detectors~\citep{abbott2016binary}.  Left undisturbed, BHs are quiet. They emit 
no GWs. However, when a BH is perturbed, say by a collision through inspiral with another BH 
~\citep{peters1963gravitational, pretorius2005evolution}, it will emit GWs at frequencies 
that can be predicted theoretically.  A deformed BH quickly dissipates energy in the form of 
GWs until it again reaches a quiescent state.  The GW emission of a perturbed BH depends 
on two factors: the BH's angular momentum and mass~\cite{MassMomentumQNM}.
The emission pattern consists of a spectrum of quasinormal modes, much like the modes 
of a bell or drum.

All musical instruments have a signature tone color dictated primarily by its shape, materials, 
and the mechanics of how it is played. These characteristics enable us to distinguish between 
different instruments by listening to them.  Below we will explore the tone-color of a BH.

\subsection{Tone-color of a black hole}

Like bells, drums, and other various instruments, it is the available frequency 
partials that contribute to an instrument's characteristic tone-color. With BHs, 
this too is the case. According to black hole perturbation theory, there are modes 
that most likely to be excited when a black hole is disturbed. In our analysis, we do not
consider which modes may be difficult to excite or suppress. Instead, we choose to remain 
naive and assume that there should be astrophysical situations where it is physically possible 
to excite any mode. Therefore, we try to show that, based on the modes that are excited, BHs 
can emulate the sound of a bell, or of a drum. Our analysis is in no way restrictive of the 
possible rich timbres that BHs could produce. We merely constrain our study to these two cases. 
The frequencies and damping times of these modes 
are only dependent on its two physical parameters - its mass and spin. Roughly speaking, the
mass determines the fundamental frequency of the BH ringing and the
spin determines the quality factor, i.e., duration, of the modes or overtones. A higher spin
allows for less damped, longer lasting modes. This is analogous to a bell, where
the quality of the bell describes its ring-time. All BHs are equivalent, assuming they have 
the same mass and spin. Other than these two factors, the composition of all BHs are the same.

Marc Kacc presented the question ``Can you hear the shape of the drum?''~\citep{hearDrum}. Through 
experience and intuition, our ear can often recognize and determine properties of a drum, including 
where it is tapped. However, since Kacc's paper, it has been proved that the answer to this question is sometimes 
negative~\cite{Mode_degen}, where there could be multiple shapes that provide the same sound, leading to 
a degeneracy. In principle, the same intuition could be developed wherein we could
hear differences among different perturbed BHs and
potentially predict their characteristics - the mass and spin. Notwithstanding, like with drums, 
it is likely that a degeneracy could occur where there are multiple modes with the same sound,
leading to an uncertainty of the BH parameters. 
Later, we show that a BHs truly sounds like unique musical instruments, precisely determined by 
which modes are excited, and by how fast it is spinning. Therefore, based on the manner of 
perturbation, a BH can be made to sound like a bell or a drum.

\begin{figure}[htpb]
\begin{center}
\includegraphics[width=\textwidth]{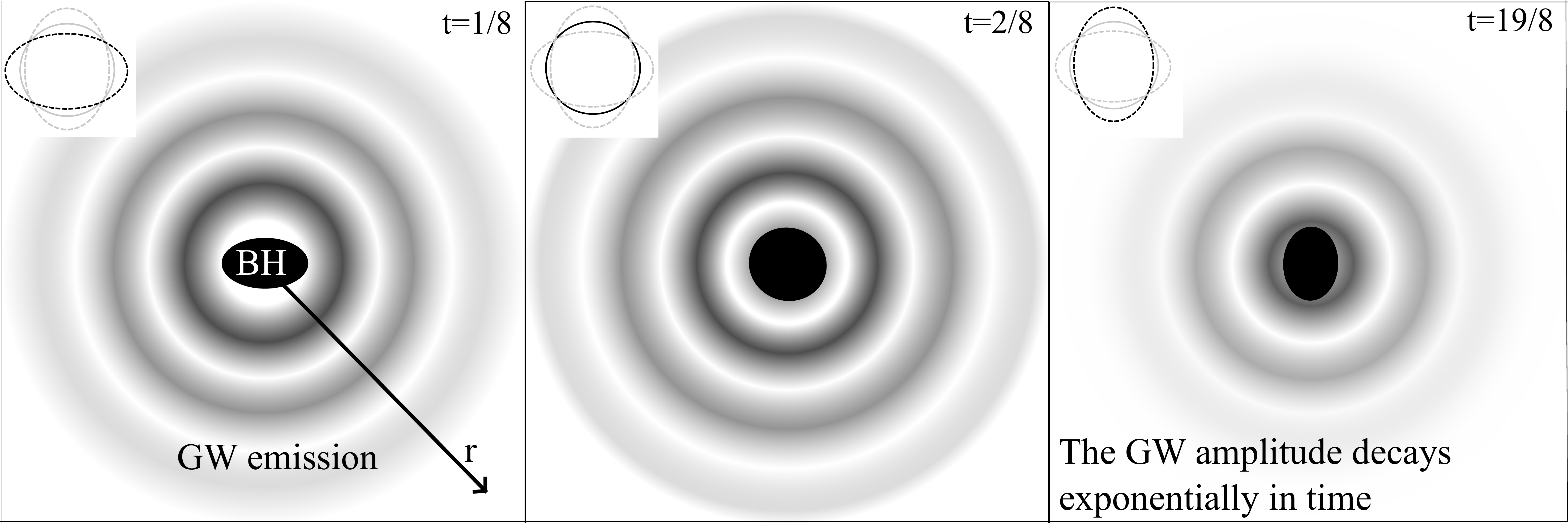}
\caption{Artist's rendition of GWs emitted from a perturbed
BH.  Shown is the equatorial plane of a BH that is ``ringing''
with an $m=2$ mode where the concentric circles indicate the GW strain with 
darker values indicating regions of a higher strain.  The three panels show the 
vibration of the BH at three distinct times. Note that the GW strain 
decreases with radial distance, $r$, from the source and with time, $t$. 
The top left insets show the relative phase of the $m=2$ mode for each panel. 
LIGO and Virgo detect the passing crests and troughs of gravitational strain 
emanating from the BH as they reach our terrestrial detectors
}
\label{fig:BHmodes}
\end{center}
\end{figure}

A perturbed BH emits GWs according to a spectrum of
possible quasinormal modes.  The spectrum is defined by the indices of
spheroidal harmonics $l,m$ that describe the BH perturbations along
with an overtone factor $n$.  Figure \ref{fig:BHmodes} shows a qualitative
diagram of an $m=2$ mode of a perturbed BH, defined by a vibration that
stretches and squeezes the space around the BH in a time-dependent way.
These waves have frequencies that depend only on the mass and spin of the BH,
and a damping time determined by the quality factor, $Q$, which depends only on
the spin parameter of the BH, $j$.

Determining the precise spectrum of BH quasinormal modes has been an
active research field for over sixty years. \cite{berti2009quasinormal}
provides a thorough and modern review on the topic and we use their fitting
formulas for the quasinormal mode frequencies and quality factors.  The
frequency $f_{lmn}$, quality factor $Q_{lmn}$ and damping time $\tau_{lmn}$ 
of a given mode indexed by the two spheroidal harmonic numbers $l$ and $m$, 
as well as an overtone number $n$, are given by
\vspace*{1pt}
\begin{align}
f_{lmn} &\approx \frac{1}{2 \pi GM/c^3} \left( \f_{1lmn} + \f_{2lmn} (1 - j)^{\f_{3lmn}} \right),
\label{eq:f}
\end{align}
\begin{align}
Q_{lmn} &\approx \frac{1}{2 \pi GM/c^3} \left( \q_{1lmn} + \q_{2lmn} (1 - j)^{\q_{3lmn}} \right),
\label{eq:Q}
\end{align}
\begin{align}
\tau_{lmn} &\equiv \frac{Q_{lmn}}{\pi f_{lmn}},
\label{eq:tau}
\end{align}
where $\f_1, \f_2, \f_3, \q_1, \q_2, \q_3$ are the given fitting parameters, unique 
to each quasinormal mode derived from GR. 
M is the mass of the BH, $j$ is the dimensionless spin parameter, which can take on
any value in the range of $[0, 1]$, $G$ is the gravitational constanti, and $c$ is the
speed of light.

\begin{figure}[htpb]
\begin{center}
\includegraphics[width=\textwidth]{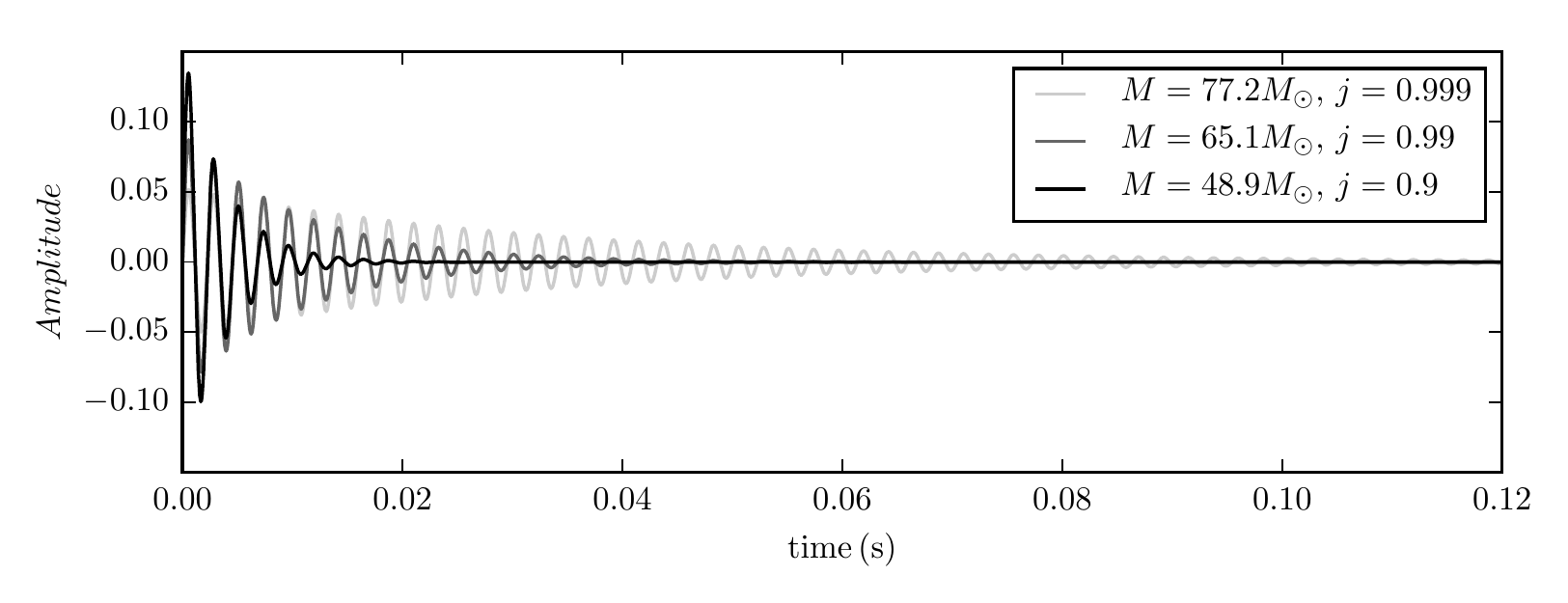}
\caption{
The $l=2,m=2,n=0$ mode of three distinct BHs with different mass and
spin.  The masses and spins are chosen such that each produces an $A_4$ (440 Hz)
as the central frequency for this mode.  In general, additional modes for each
BH may be excited according to \eqref{eq:modes}. 
}
\label{fig:waveform}
\end{center}
\end{figure}

Neglecting geometric factors, the detected GW strain from a
perturbed BH at a detector on Earth is given by
\begin{align} 
h_{lmn}(t) &= e^{-t / \tau_{lmn}} 
	   \left[ A_{lmn} \sin(2\pi f_{lmn} t)
	   + B_{lmn} \cos(2\pi f_{lmn} t) \right],
\end{align}
where $A_{lmn}$ and $B_{lmn}$ correspond to complex mode amplitudes. 
In this work, we will only consider the sine phase to simplify the discussion
and note that it has the benefit of being zero at time zero. Furthermore, we
will consider modes up to $l=m=4$ and $n=3$, for which the fitting formulas are
defined in~\cite{berti2009quasinormal}. Therefore, the waveforms we consider are linear
combinations of the sine phases of the above modes given by
\begin{align} 
h(t) &= \sum_{n=1}^3\sum_{l=2}^4\sum_{m=-l}^l 
     A_{lmn} e^{-t / \tau_{lmn}} 
     \sin(2\pi f_{lmn} t).
     \label{eq:modes}
\end{align}
The amplitude of the mode, $A_{lmn}$ encodes how the BH is perturbed
and we leave it as a free parameter. We show that it is possible to obtain the $A_4$ 
frequency from a variety of black holes for the same mode in Figure ~\ref{fig:waveform}
below

\subsection{Black holes as bells or drums}

\begin{figure}[htpb]
\begin{center}
\includegraphics[width=\textwidth]{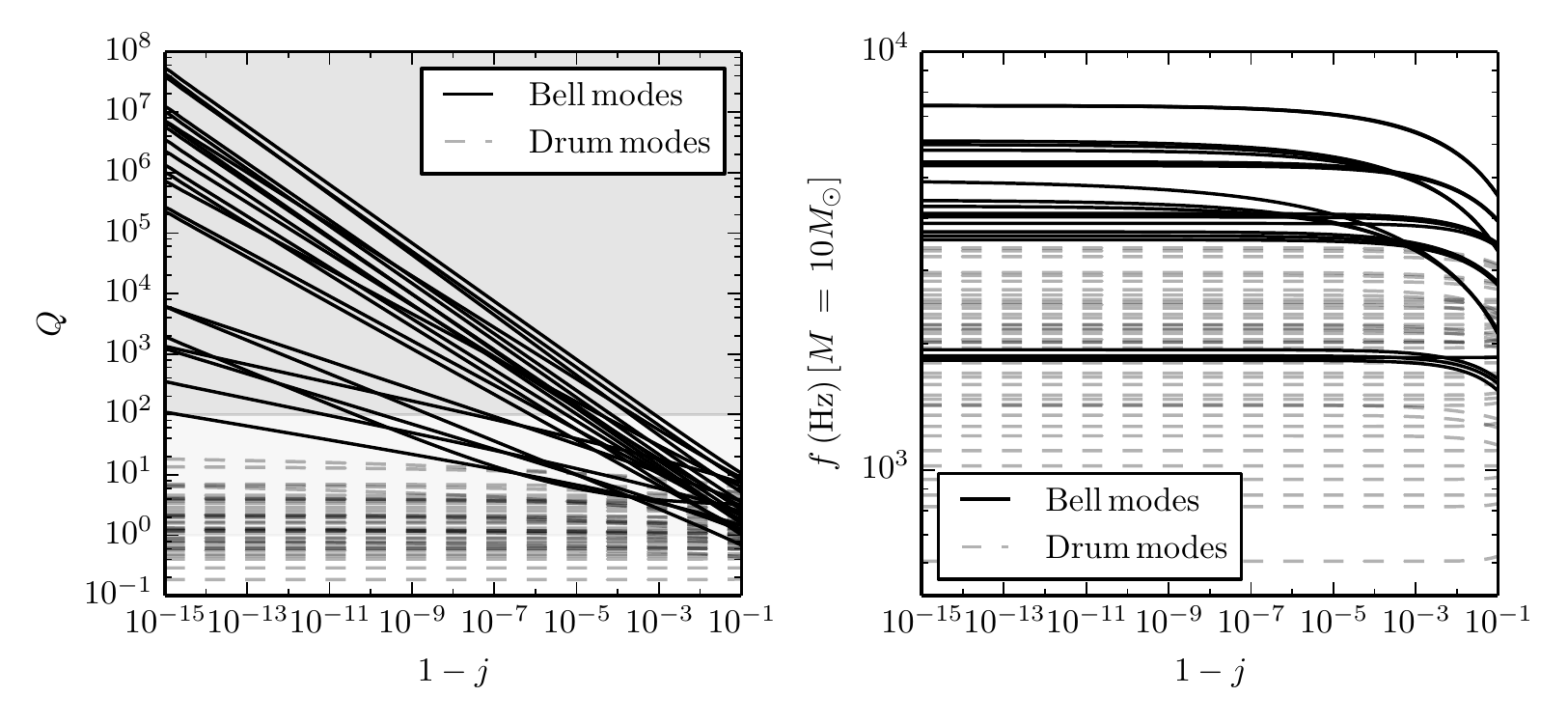}
\caption{
Quality factor, $Q$, (left) and frequency, $f$, (right) for the first 63 modes
of a BH defined by \eqref{eq:Q}. As $1-j$ approaches zero, $Q$ diverges
for modes with $q_3 < 0$, but $f$ asymptotes to a constant value. We denote
modes with $q_3 < 0$ as ``bell-like'', since as $j$ approaches one the Q values
exceed 100.  The remaining modes we denote as ``drum-like''.  As $j$ approaches
zero, all the modes are drum-like.}
\label{fig:modes}
\end{center}
\end{figure}

A bell's ring typically lasts for many cycles of the fundamental frequency,
i.e., a large $Q$, whereas a drum's duration could be very short, i.e., a small
$Q$. From experience, we have seen that too low a quality factor, i.e., $Q \lesssim 5$ leads 
to more of a click, rather than drum-thud type sound. We will classify BHs that 
produce long-lived gravitational-wave ringing as bell-like, and those with short 
ringing as drum-like/click-like. To distinguish the two classes, we will arbitrarily establish
$Q=100$ as the threshold between a drum/click and a bell. Subsequently, we will establish 
$5 \leq Q \leq 100$ as drum-like modes to avoid click-type tones. The fitting parameters $\q_1,
\q_2, \q_3$ range from $\sim [-20, 20]$.  In order to have large $Q$, e.g.,
$Q>100$, two conditions are required: 1) $\q_3 < 0$ and 2) $j \sim 1$.  
These bell-like and drum-like modes are separated into two categories as seen in
Figure~\ref{fig:modes}.

\subsection{Can a black hole make a good bell or drum?}
\label{sec:goodbell}

Black holes, under various manners of perturbations, can form quite rich and unique sounding timbres.
Since there is a myriad of various unique timbres yet to be found, we encourage the reader to 
use our software to come up with new timbres that have not yet been characterized or emulated 
by current, man-made musical instruments. However, for our analysis, we intend to carry out a simple 
analysis that compares bells and drums with their respective BH approximations. 

\begin{figure}[htpb]
\begin{center}
\includegraphics[width=\textwidth]{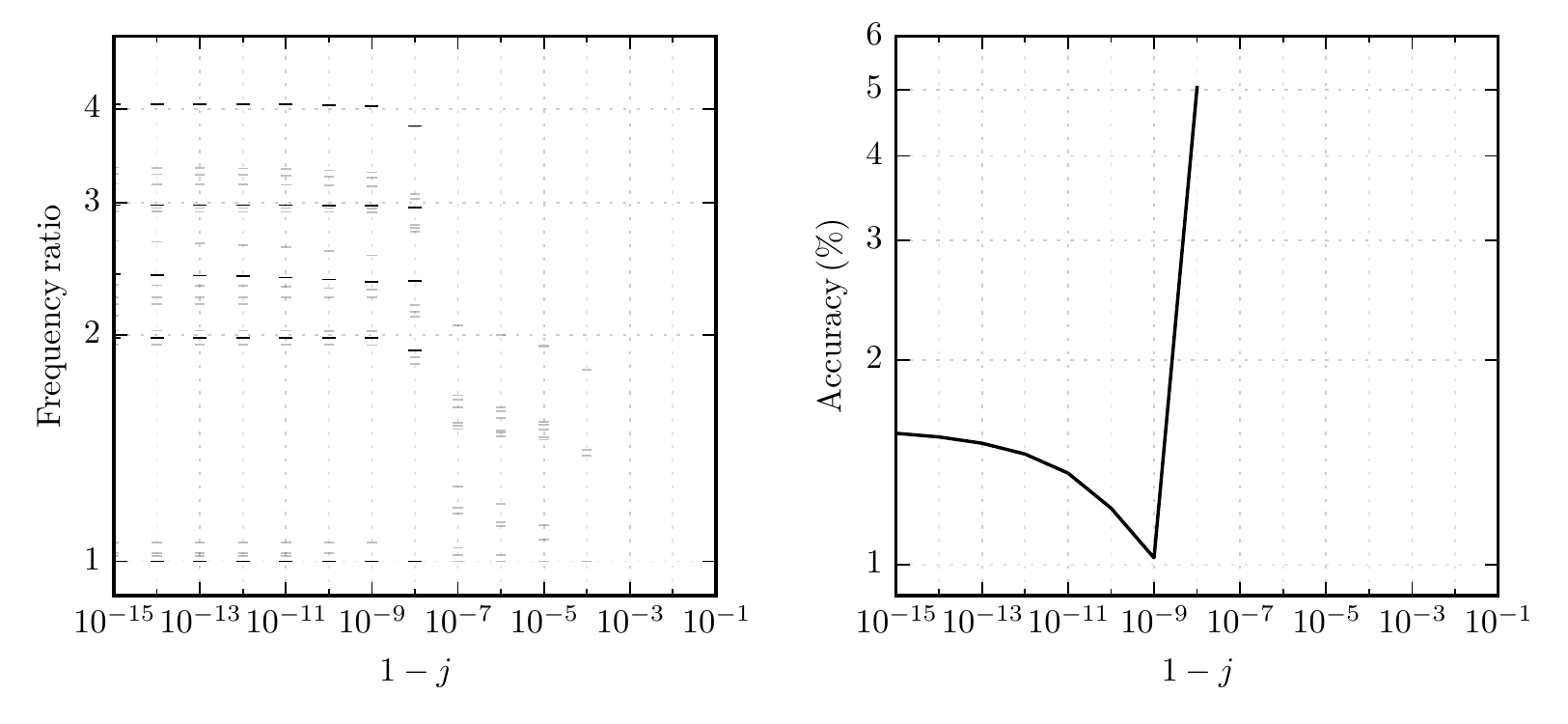}
\caption{
Frequency spacing of the bell-like modes of a BH (left) and accuracy to
which a ``typical good quality church bell'' can be achieved (right) for the
worst mode.  We restricted the modes in this figure to be those with a quality
factor, $5 \leq Q \leq 100$.  Under these constraints, it can be seen that a
quality bell constrained by the nominal frequency ratio ($1:2:2^{15/12}:2^{19/12}:4$, relative to the hum frequency) is only achieved when $1-j < 10^{-9}$.  
The damping timescales are considerably different for each mode as indicated by Figure 
\ref{fig:modes}. This implies that although a BH may start out sounding like a church
bell, the modes will not decay as one would expect from a typical bell.
}
\label{fig:perfectBell}
\end{center}
\end{figure}

Perrin et al~\cite{PERRIN198329} note that a ``typical good quality church bell'' in western
music has its first five partials with the frequency ratios of $1:2:2^{15/12}:2^{19/12}:4$. It is important to note that this is not the only prescription 
for a church bell. However, for our analysis, we try to match a BH to this particular 
frequency ratio. 
The modes of BHs do not match these ratios, as the ratios fall between these
values. Instead, we determine whether or not the ratios between \textit{any} of
the first 63 modes defined by Equation \eqref{eq:modes} for a given BH spin to 
have similarities to the spectrum of a bell. Figure \ref{fig:perfectBell} shows the
frequency spacing of the bell-like modes, $\q_3 < 0$, as a function of $1-j$, as
well as the accuracy with which a nominal bell can be obtained. With a one percent 
accuracy, a BH with a spin greater than 0.9999999 can mimic an ideal
bell (with Perrin's prescription) in terms of frequency content.  However, each mode has a 
substantially different quality factor, $Q$, meaning that the partials do not decay at a similar 
rate.  So, although a perturbed BH may initially resemble the spectra of a typical
good quality bell, the modes will not decay in the same way. We explore more of this 
particular example later. 

\begin{figure}[htpb]
\begin{center}
\includegraphics[width=\textwidth]{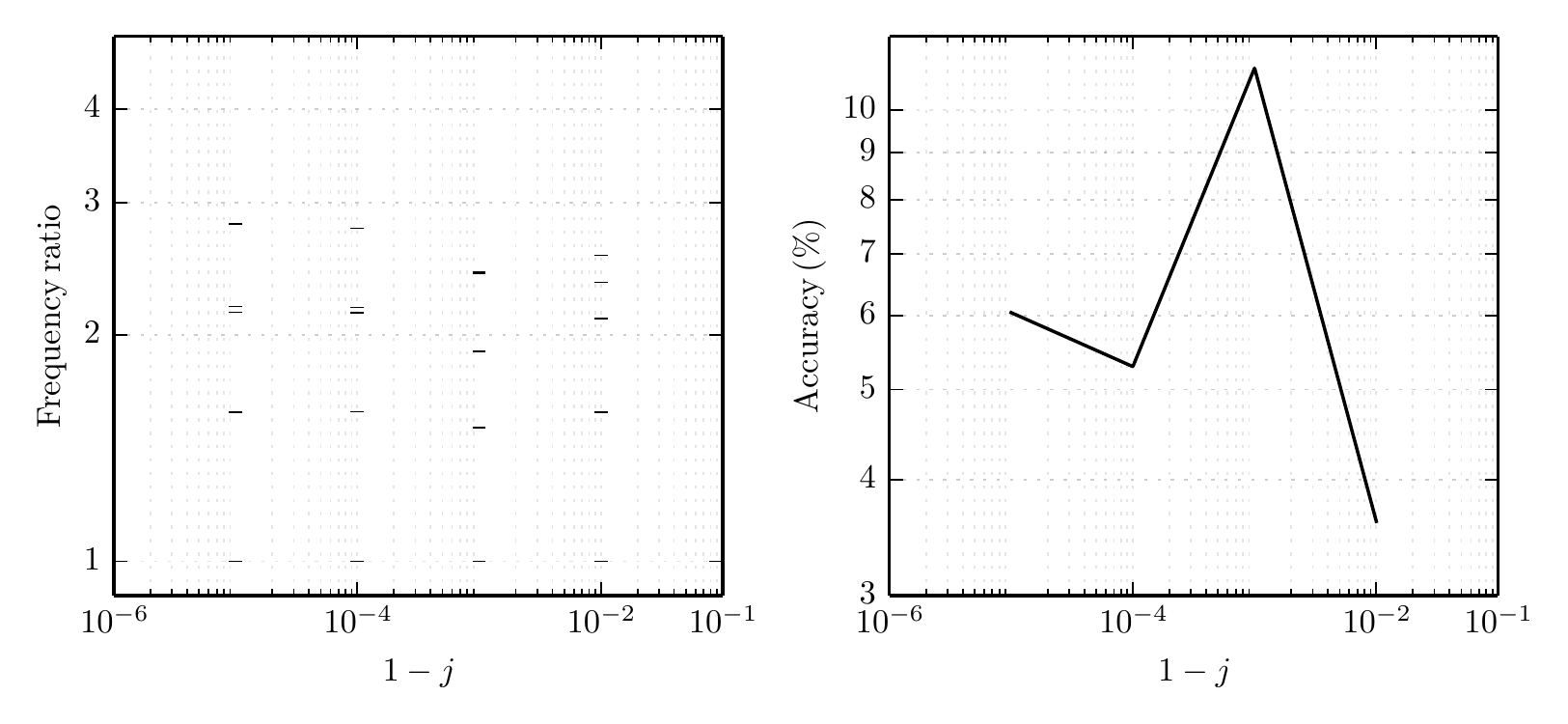}
\caption{
Frequency spacing of the drum-like modes of a BH (left) and accuracy to
which a ``typical good quality drum" can be achieved (right) for the 
worst mode.  We restricted the modes in this figure to be those with a quality
factor $5 \leq Q \leq 100$
Under these constraints, it can be seen that a quality drum constrained by the nominal 
frequency ratio ($1:1.59:2.14:2.30:2.65$, relative to the first overtone) is only 
achieved when $10^{-5} \leq 1-j \leq 10^{-2}$. The damping timescales are considerably 
different for each mode as indicated by Figure \ref{fig:modes}. This implies that 
although a BH may start out sounding like a drum, the modes will not decay 
as one would expect from a typical drum.  
}
\label{fig:perfectDrum}
\end{center}
\end{figure}

Black holes may also be compared to a simple drum membrane. We repeat the same 
process, this time matching \textit{any} black hole mode frequencies to have ratios 
representative of drums: $1:1.59:2.14:2.30:2.65$, relative to the first overtone~\cite{drumPartials}. 
Just as in the bell case, it is important to note that this 
is not the only prescription of overtone-ratios for a drum. However, for our analysis, we 
try to match a BH to this particular frequency ratio.
Figure \ref{fig:perfectDrum} shows the frequency spacing of the drum-like modes, as a function of $1-j$, as
well as the accuracy with which an ideal drum can be obtained. We have placed constraints
on the drum like modes to have $5 \leq Q \leq 100$, to a) exclude bell-like modes and 
b) avoid click-like sounds. Under these constraints, we see that spins values of 
$10^{-5} \leq 1-j \leq 10^{-2}$ are needed to obtain values close to this nominal frequency 
ratio. Later on, we explore an drum membrane example subject to these constraints.
An important caveat to note is that although high spins are required to get accurate partials 
for the listed nominal bell and drum frequency ratios, the errors associated with the fitting 
coefficients in this part of the parameter space have not been considered. More recent work have 
found more accurate methods in computing these factors in the extremal BH regime~\cite{extremal_fits}. 
Moreover, for spins crossing the Thorne limit of $j = 0.998$,
the resulting BH is likely not astrophysical~\cite{ThorneLimit}.
 
For both the bell and drum cases, we also plot the accuracy 
of the mode with the largest deviation from a nominal frequency ratio. This 
estimation in accuracy does not take into account the errors associated with the 
fitting factors listed in Equation \ref{eq:f} (see \cite{berti2009quasinormal} for a 
complete list of tabulated errors), and is computed as follows:
\begin{align}
Accuracy = \max_i\left[\min_{lmn}\left(\frac{f_{lmn}}{f_{\text{partial}_i}}\right)\right]
           \times 100.
\end{align}

\section{Carillons and drums from black holes}

We have shown that for particular spins, it is possible to create black hole modes that 
approximate bell-like and drum-like timbres, derived from emitted GWs.
While it may be possible to create instruments with more complicated waveforms, 
through more complicated astrophysical scenarios, we limit our study to these 
simpler instruments. Using the algorithm described above to find five accurate partials,
we create an approximation to a church bell and a simple drumhead.

Our example instrument tones are tuned by setting the prime frequency to be the 
fundamental pitch. 
The 88 bells, each assigned to a chromatic pitch, create a kind of digital carillon:
an instrument that contains an array of large church bells, played by striking a lever 
connected to the keys of a special type of piano. As there is much study done on the 
modal structure of a historic carillon in the city of Flanders, Belgium, we can 
compare our BH carillon to this historic carillon. 

\subsection{Typical church bell}

\begin{figure}[!htpb]
\begin{center}
\includegraphics[width=\textwidth]{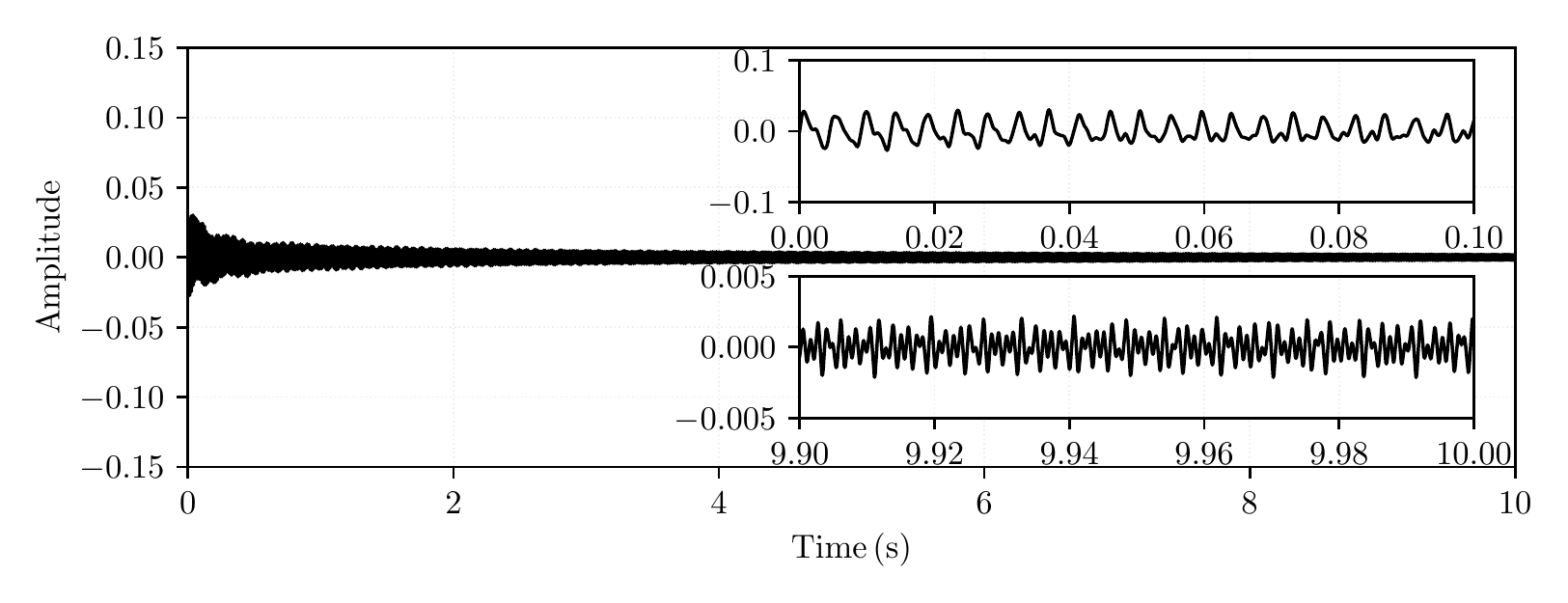}
\caption{
First 10 seconds of the waveform of good quality typical church bell tuned to
$A_4$ made from a BH.  As you can see, the waveform is very long-lived
$>10$ s and evolves over time as indicated by the difference between the top
inset (first 0.10 seconds) and bottom inset (last 0.10 seconds).
}
\label{fig:bh_bell_wav}
\end{center}
\end{figure}

As described above, it is possible to approximate a church bell-like sound
based on a BH, although, the result differs from typical bell spectra due to 1) the 
ratios of mode frequencies not exactly matching those of a bell and 2) relative mode 
decay times sustaining longer than those of a bell. 
Table \ref{table:bh_bell} shows a simple set of modes. It can be seen that the decay 
times increase with higher partials, which leads to a change in the waveform at later times 
(see Figure~\ref{fig:bh_bell_wav}). This is in direct contrast to the behavior of 
bell partials, which typically feature the longest sustain from the first partial (hum), 
and quicker decay times in the higher partials. 

\begin{table}[h]
\centering
\begin{tabular}{c c c c}
\textbf{partial} & \textbf{mode} & \textbf{f[Hz]} & \textbf{Q} \\
\hline 
Hum         & (2, 1, 2) & 222 & 131  \\
Fundamental & (3, 2, 1) & 440 & 1660 \\
Minor third & (2, 2, 1) & 522 & 9840 \\
Fifth       & (4, 3, 0) & 661 & 19600\\
Octave      & (4, 4, 0) & 896 & 68600
\end{tabular}
\caption{
A BH approximation to a good quality typical church bell tuned to $A_4$.  Here the
mass is $M = 82.37 M_\odot$ and the spin is $j = 0.999999999$. The tuning is fixed by 
setting the $l=3,m=2,n=1$ mode to be the fundamental frequency, with the largest 
amplitude.
}
\label{table:bh_bell}
\end{table}

\subsection{Comparison to the actual carillon}

This set of partials and decay times can be compared to the bells of a carillon cast by 
Joris Dumery ca. 1742~\cite{CarillonBell}. The modal structure of the Dumery bells has 
been studied extensively, particularly their frequency dependence on mass and diameter. 
Of particular interest are the first five partials, which we compare to the BH 
bell's modal dependence on mass and geometry. 
 
As mentioned before, the modes of a BH depend on a 3-dimensional spheroidal 
geometry. For bell-like structures, modes are described as nodal meridians $m$,
and nodal circles $n$ (see \cite{strutt_2011} for a reissue of Lord Rayleigh's work). Since there is no obvious 
translation between these two geometries, we list the first five Dumery bell partials 
in Table \ref{table:bh_carillon} which result from full meridians, as seen in \cite{CarillonBell}.

\begin{table}[h]
\centering
\begin{tabular}{c c c}
\textbf{partial} & \textbf{BH mode} & \textbf{carillon mode (m,n)} \\
\hline 
Hum         & (2, 1, 2) & (2, 0)  \\  
Fundamental & (3, 2, 1) & (2, 1)  \\
Minor third & (2, 2, 1) & (3, 1)  \\
Fifth       & (4, 3, 0) & (3, 1)  \\
Octave      & (4, 4, 0) & (4, 1) 
\end{tabular}
\caption{
Comparison of the modal structure of first five partials of an ideal BH
bell with the first five partials of a carillon bell. m and n for bells stand 
for nodal meridians, and nodal circles respectively.
}
\label{table:bh_carillon}
\end{table}

\begin{figure}[!htpb]
\begin{center}
\includegraphics[width=\textwidth]{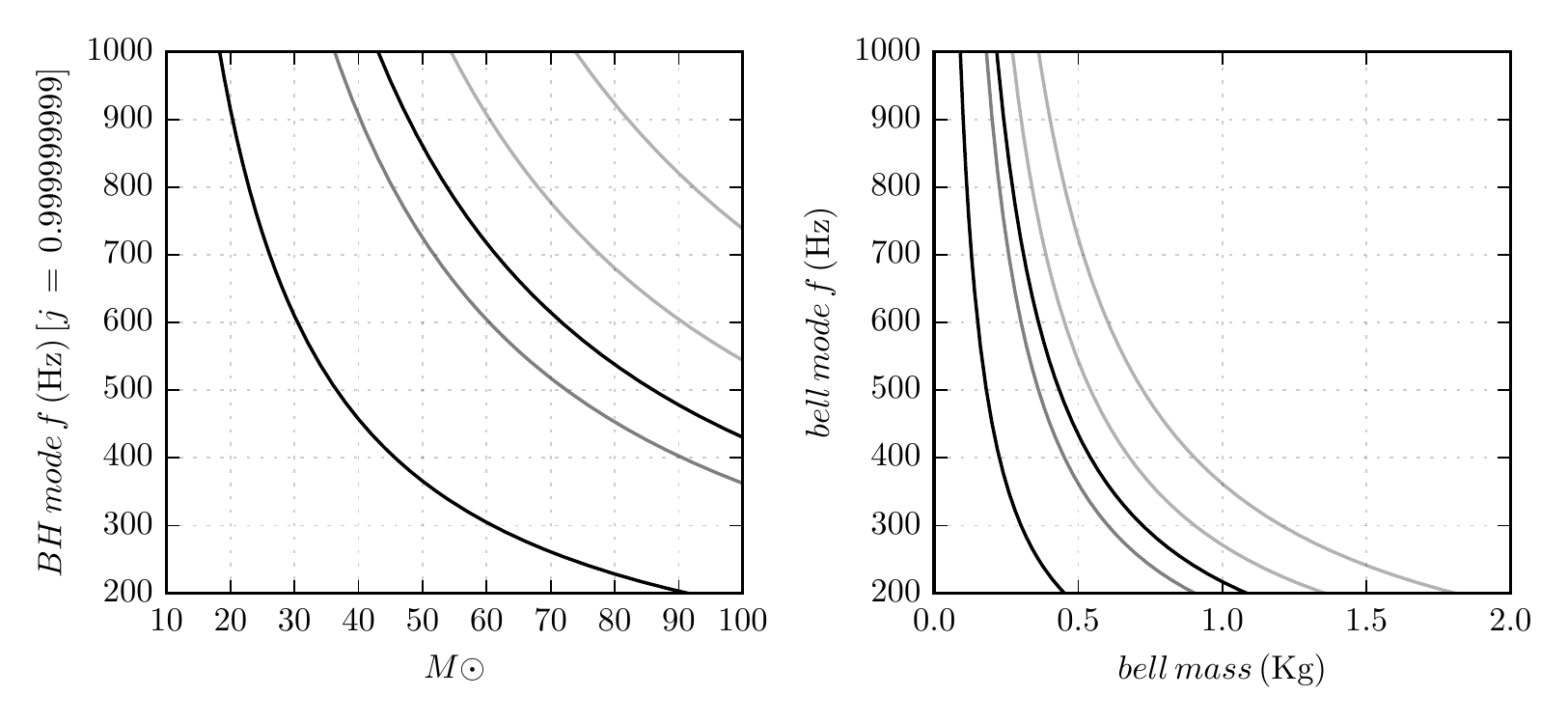}
\caption{
Left figure: Frequency dependence of quasinormal modes on BH mass $(M_\odot)$,
changed by varying spin. From left to right: BH partials are hum, minor third, 
fundamental, fifth, octave.
Right figure: Frequency dependence of bell modes on bell mass~\cite{CarillonBell}.
From left to right: bell partials are hum, fundamental, minor third, fifth, octave.
}
\label{fig:BH_bell_mass}
\end{center}
\end{figure}

We first investigate mode frequency dependence on mass for BHs and carillon bells, 
and then consider frequency dependence on geometry. By fixing the spin of an array of 
BHs with varying mass, we can utilize the methods of Equation \eqref{eq:f} to compute BH mode 
frequency dependence on mass. In Figure \ref{fig:BH_bell_mass} we present the results 
alongside carillon bell mode frequency dependence on bell mass, computed from~\cite{CarillonBell}.

While it was relatively easy to compare mode frequency dependence on mass for bells and 
BHs, the problem becomes less well-defined when computing the mode frequency 
dependence on a BH's diameter. To quantify a diameter, we seek to assign the 
radius from which GWs are emitted. From GR, it follows that for spinning 
BHs, an axisymmetric surface is created from which GWs will propagate, without 
falling in. While this surface is spherical for a static BH, due to the spin it morphs into an 
axisymmetric surface with two radii that encompass a region termed the light-ring. 
To simplify further discussion, we consider only one coordinate radius, $r$, of this light-ring~\cite{KerrLightCurve}
\vspace*{1.5pt}
\begin{align}
r &= \frac{2MG}{c^2} \bigg[1 + \, \cos\bigg(\frac{2}{3}\cos^{-1}(-j)\bigg)\bigg], 
\label{eq:kerrRadius}
\end{align}
\vspace*{1.5pt}
where $M$ is the mass of the BH, and $j$ is the dimensionless spin parameter. Given this 
BH 'diameter' $(D = 2r)$, we can begin to investigate 
the frequency dependence of BH modes based on its diameter, and examine how it differs
with the carillon bells' mode frequency dependence on bell diameter.
 
\begin{figure}[!htpb]
\begin{center}
\includegraphics[width=\textwidth]{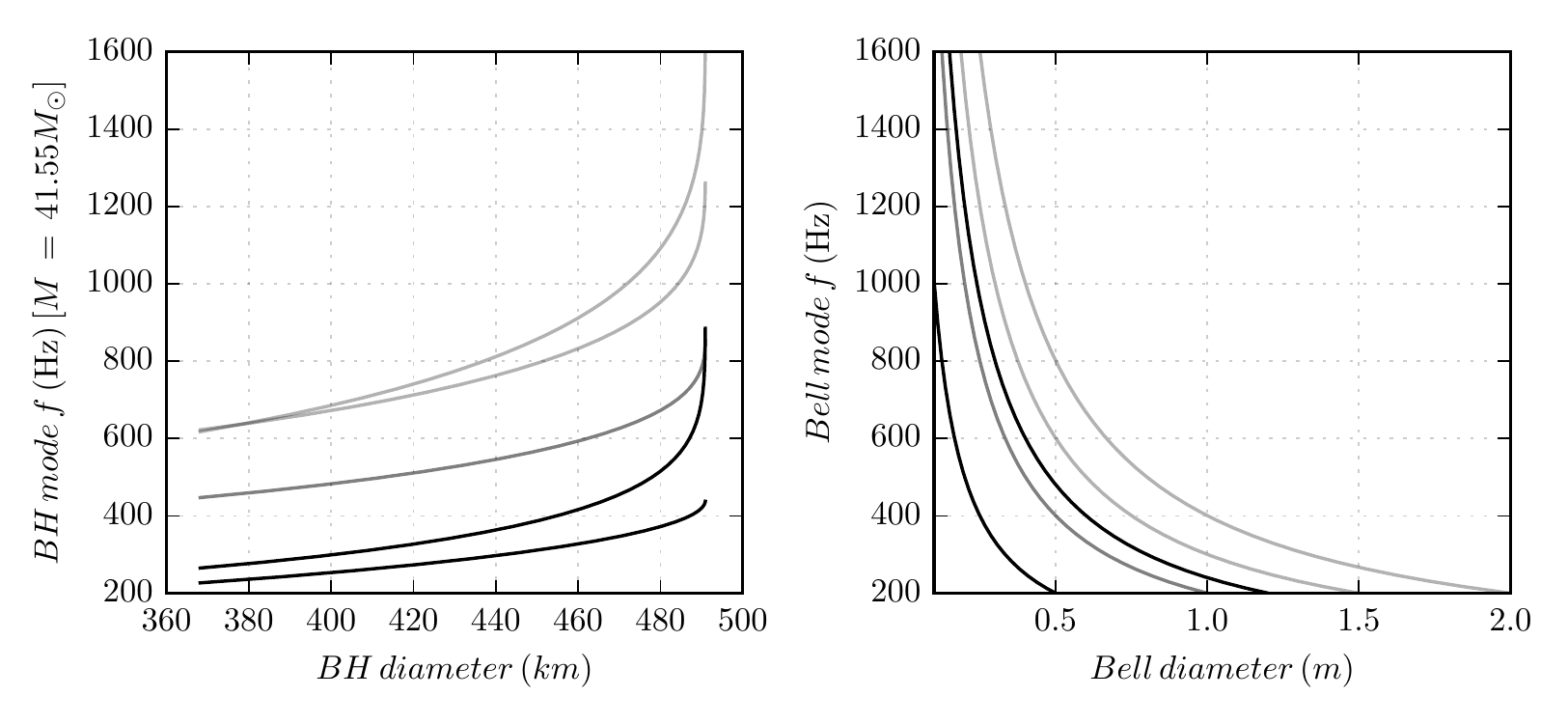}
\caption{
Left figure: Frequency dependence of quasinormal modes on BH diameter $d$, 
changed by varying spin. From top to bottom: BH partials are octave, fifth, fundamental, 
minor third, hum.
Right figure: Frequency dependence of bell modes on diameter~\cite{CarillonBell}. 
From left to right: bell partials are hum, fundamental, minor third, fifth, octave.  
}
\label{fig:BH_bell_diam}
\end{center}
\end{figure}

Only a change in the mass or spin of a BH can change the diameter of its light ring.
By choosing a fixed mass, we can see how the radius changes with spin. We can reframe 
Equation \eqref{eq:f} using Equation \eqref{eq:kerrRadius} to change the mode frequency dependence on 
spin $j$ to its diameter $D$. Our results are plotted in Figure \ref{fig:BH_bell_diam} and 
compared to Dumery bell mode frequency dependence on bell diameter~\cite{CarillonBell}

\begin{align}
f_{lmn} &\approx \frac{1}{2 \pi GM/c^3} \left[\f_{1lmn} + \f_{2lmn} \bigg[1 + \, \cos\bigg(\frac{3}{2}
        \cos^{-1}\bigg(\frac{rc^2}{2MG} - 1\bigg)\bigg]^{\f_{3lmn}} \right].
\label{eq:f_diam}
\end{align}

\subsection{Typical drum membrane}

\begin{figure}[!htpb]
\begin{center}
\includegraphics[width=\textwidth]{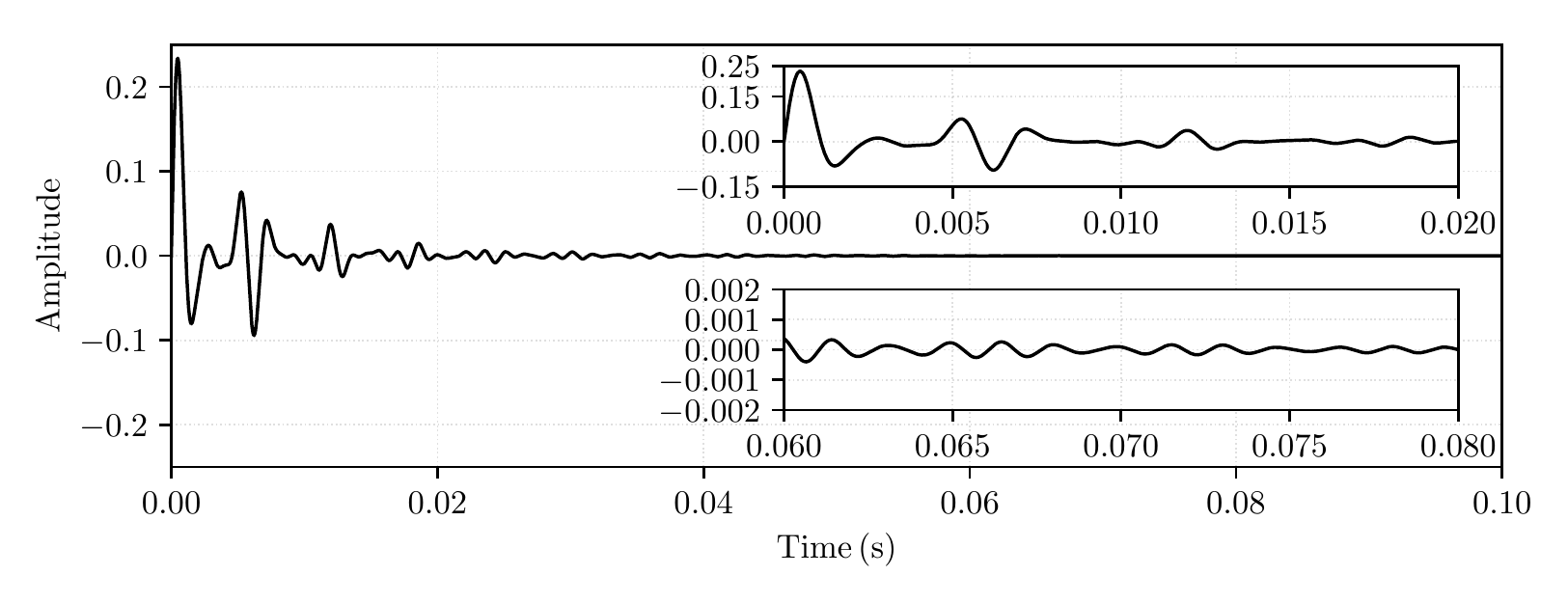}
\caption{
Waveform of a drumhead membrane simulated with five partials tuned to $A_4$ made from a BH. 
As you can see, the waveform is not long-lived ($~0.3$ s) and evolves over time as indicated by 
the difference between the top inset (first 0.02 seconds) and bottom inset (later 0.02 seconds). 
}
\label{fig:bh_drum_wav}
\end{center}
\end{figure}

For BHs with comparatively low spin, the quick decays of the modal frequencies emulate the thud 
of a typical drum membrane. Here, we present a very simple drum tuned to $A_4$ created by a BH 
spinning less rapidly than in our previous example $(j = 0.9999)$, created from the overtones listed below. 
For quality factors $Q \lesssim 5$, we found that our approximation resembled a click, 
rather than the characteristic thud of a drum, while values of $Q > 100$ produced 
bell-type ringing modes. From Figure~\ref{fig:perfectDrum}, it can be seen that there 
are fewer mode combinations that satisfy these constraints. We have listed the one 
with an accuracy of five percent in Table~\ref{table:bh_drum}. As in the bell approximation, 
the decay times increase with higher partials for our drum membrane approximation. Although the 
decay times are not as widely spread in this case, and the waveform is not long lived, there still 
occurs a change in the waveform at later times (see Figure \ref{fig:bh_drum_wav}).

\begin{table}[h!]
\centering
\begin{tabular}{c c c}
\textbf{mode} & \textbf{f[Hz]} & \textbf{Q} \\
\hline 
(2, 1, 1) & 278 & 6  \\ 
(4, 0, 0) & 440 & 6  \\
(4, 2, 1) & 596 & 9  \\
(4, 2, 0) & 606 & 27 \\
(4, 3, 2) & 771 & 29 
\end{tabular}
\caption{
A BH approximation to a good quality drum tuned to $A_4$.  Here the mass is $M = 67.03 M_\odot$ 
and the spin is $j = 0.9999$. The tuning is fixed by setting the $l=4,m=0,n=0$ mode to be the 
fundamental frequency, with the largest amplitude.
}
\label{table:bh_drum}
\end{table}

\section{Constructing a black hole instrument}

In order to construct a BH instrument, we create 88 bell-like timbres, save them
as audio files, and assign each to a pitch that can be played by a MIDI keyboard. 
To start the process we construct our canonical BH by choosing a
canonical mass $M_{\mathrm{can}}$ and a user-defined set of mode amplitudes,
$A_{lmn}$.  We denote the primary tone as the mode frequency,
$f_{\mathrm{can}}$ with the largest $A_{lmn}$.  From our canonical BH
we rescale the canonical mass to match the 88 keys of the piano through
\begin{align}
\frac{M}{M_{\mathrm{can}}} &= \frac{f_{\mathrm{can}}}{f_k},
\end{align} 
where $f_k$ is given by
\begin{align}
f_k &= 2^{(k-49) / 12} \times 440 \,\text{Hz},
\end{align}
where $k$ denotes the keys of the keyboard (from 1 to 88).

\section{Conclusion}

We have showcased our method of creating digital BH instruments, and hope that this 
project inspires researchers and the public into creating their own types of instruments.
Furthermore, we would like to highly encourage the reader to find BH timbres that do not 
fit the mould of current musical instruments.
The music that we create is from the perturbations of BHs, but understanding what kind 
of exact perturbation causes a BH to sound like a bell, drum or something else 
is out of the scope of this paper. We have seen that from the literature that the impact time 
of a force imparted to a bell directly contributes to which higher modes are excited, which in 
turn decides the tone-color of the sound emitted~\cite{bellHigherModes}. We predict that, 
in turn, the manner and duration of a BH perturbation directly lends to the tone-color of the emitted GW.  

One major limitation of our BH approximation of musical instruments is that the black 
hole instruments are only created from five partials. Therefore, at most one can approximate an 
ideal, generic bell. To be able to emulate a particular bell, more higher 
partials are needed to capture its true tone-color. Nevertheless, even if there were fitting 
factors that could derive frequencies of higher partials for any BH mass or spin, 
the accuracy would undoubtedly plummet as there are only so many degrees of freedom available 
in choosing accurate partials. A possible solution to this is to have more overtones, i.e.,  
$n > 3$ to have a larger pool of modes to choose from. 

Another point to consider is the tuning of our BH bells and drums. We have simplified 
our analysis by setting the second partial (prime) to be the fundamental frequency, and having 
the largest amplitude, while this is not the case with all bells. We have also ignored 
issues of the perceived pitch at the beginning of perturbation. Moreover, we have found an inverse 
structure to the quality factor, $Q$, of our bell and drum approximations. In bells, the 
lower partials have a higher quality factor, with the hum frequency being longer lived than 
the higher partials. We find that this is not the case with BH bells. There seems to be 
a transfer of energy from lower to higher partials, which results in the tuning of the bell 
approximation being fundamentally changed at later times. 

Black hole perturbation theory is assisting GW astronomers to understand the 
parameters of observed BHs by analyzing the strain waveforms. We hope that this 
exploration into the similarities of modal structure between BHs and bells/drums, 
and the existing studies done on mode frequency dependence on the features of bells will assist
the scientific community to obtain a different avenue in which to further explore the structure of 
BHs.

\section{Acknowledgments}

This work would not have been possible without the guidance and assistance of the Penn State LIGO team, 
especially Bangalore S. Sathyaprakash, Cody Messick and Alexander E. Pace.
We thank Ra Inta and Dani P. Rajan for useful discussions. D.G., D.M. and C.H. also thank M.B. and 
Lauren Edwards for using one of our black hole instruments to re-create the Star Trek theme song.
Finally, this manuscript was improved by comments made by Ra Inta and Nathan Johnson-McDaniel.
This work was supported in part by the National Science Foundation through PHY-1454389. 
This document has LIGO preprint number P1800019-v3.

\bibliographystyle{plain}
\bibliography{cmjbib}

\end{document}